# Multi-Color Spaceplates in the Visible

Masoud Pahlevaninezhad[1] and Francesco Monticone[1,*]

[1]School of Electrical and Computer Engineering, Cornell University, Ithaca, NY, USA.

*Corresponding author. Email: francesco.monticone@cornell.edu

**Abstract**

The ultimate miniaturization of any optical system relies on the reduction or removal of free-space gaps between optical elements. Recently, nonlocal flat optic components named "spaceplates" were introduced to effectively compress space for light propagation. However, space compression over the visible spectrum remains beyond the reach of current spaceplate designs due to their inherently limited operating bandwidth and functional inefficiencies in the visible range. Here, we introduce "multi-color" spaceplates performing achromatic space compression at three distinct color channels across the visible spectrum to markedly miniaturize color imaging systems. In this approach, we first design monochromatic spaceplates with high compression factors and high transmission amplitudes at visible wavelengths based on a scalable structure and dielectric materials widely used in the fabrication of meta-optical components. We then show that the dispersion-engineered combination of monochromatic spaceplates with suitably designed transmission responses forms multi-color spaceplates that function achromatically. The proposed multi-color spaceplates, composed of amorphous titanium dioxide and silicon dioxide layers, efficiently replace free-space volumes with compression ratios as high as 4.6, beyond what would be achievable by a continuously broadband spaceplate made of the same materials. Our strategy for designing monochromatic and multi-color spaceplates along with the presented results show that strong space-compression effects can be achieved in the visible range, and may ultimately enable a new generation of ultra-thin optical devices for various applications.

The miniaturization of optical systems has been one of the most important goals in the field of optics in recent years. The desire for compact and lightweight virtual and augmented reality (VR and AR) headsets[1] or ultra-thin cameras[2,3], for example, is posing new design challenges that are difficult to meet with standard solutions. While the recent adoption of flat optical components, known as metasurfaces[4–11], has enabled some degree of miniaturization, the main obstacle lies in the large empty spaces (free-space) between optical components, imaging sensors, displays, etc. Replacing optical elements with thin metasurfaces, while promising, often only yields incremental size reductions. Hence, the true miniaturization of an entire optical system hinges on the compression or complete removal of free-space between each of its optical/optoelectronic elements.

Free-space is an essential element in a typical optical system. For example, in a basic imaging setup, propagation of light in the space between the lens and imaging sensor is crucial to obtain the required angle-dependent phase delay and form a focused image. The idea of compressing a free-space volume into a thinner optical element, known as a "spaceplate", was recently introduced[12–18]. In particular, "nonlocal metasurfaces"[19–26] that impart an angle-dependent phase onto the transmitted light have been at the basis of most spaceplate designs proposed to date. By engineering the nonlocal (i.e., wavevector-dependent) response of these metasurfaces, propagation of light through a spaceplate can emulate the phase characteristics of free-space propagation for a distance much longer than the physical thickness of the spaceplate itself. Such a device can then be incorporated into an optical system to achieve a drastic reduction of its overall length and potentially reach its ultimate thickness limit dictated by its optical function[2,27].

In the recent literature, nonlocal metasurfaces such as photonic crystals[12,16], stacks of thin films[13,15], and coupled Fabry-Pérot (FP) resonant cavities[14,17] were engineered to achieve space compression for operation mainly in the infrared and microwave spectral regions and over narrow bandwidths. Conversely, broadband or multi-band space compression, especially at visible wavelengths, where a myriad of optical devices operate, still falls beyond the reach of current spaceplate designs. Indeed, miniaturization of many optical devices for visible light would require spaceplates that can ideally perform space compression over the entire visible range. However, a major limitation of current spaceplate designs is their inherently narrow-band response, leading to space-compression effects that are strongly wavelength-dependent. Additionally, current spaceplates rely on materials with high refractive indices such as Si and GaAs, which are highly inefficient in the visible range due to high dissipative losses.

The operating bandwidth of a spaceplate is constrained by strict physical bounds underlying generic space compression mechanisms, as shown in Ref.[18]. The fundamental trade-off between compression ratio and bandwidth, inherent in the optics of spaceplates, suggests that drastic space compression over the entire visible range is fundamentally challenging and far from feasible in practice, as it would require transparent materials with a large refractive index (e.g., a refractive index on the order of 3 to achieve a compression ratio of 4)[18]. These physical bounds, however, only apply to spaceplates operating over *continuous* bandwidths, suggesting a possible strategy to bypass these issues. Based on this insight, in this work we introduce "multi-color" spaceplates operating at three distinct color channels over the visible spectrum to replace and compress free-space in color imaging systems. The space compression effect is based on the guided-mode resonances in dispersion-engineered coupled planar FP cavities made of amorphous $TiO_2$ and $SiO_2$ layers, which are widely used in the fabrication of metasurfaces. The proposed multi-color spaceplates have high transmission efficiency and an achromatic response at three wavelengths with space compression ratios as high as ~ 4.6, using dielectric materials with a refractive index less than ~2.6. In the following sections, we first present a strategy to design monochromatic spaceplates with high transmission amplitudes and high compression factors at any visible wavelength. These monochromatic spaceplates are the building blocks of the proposed multi-color spaceplates. Next, we demonstrate that such monochromatic spaceplates with tailored transmission responses can be coupled in a dispersion-engineered manner to create multi-color spaceplates that *achromatically* compress free-space at three visible wavelengths while maintaining high performance. Trivial combinations/cascades of monochromatic spaceplates would instead lead to drastic aberrations. The presented theoretical and computational results confirm the high performance of the proposed spaceplates over the visible spectrum, in terms of transmission efficiency, usable angular range, and compression ratio. These findings may pave the way for the experimental demonstration of strong space-compression effects for visible light, a goal that has so far remained elusive.

**Results**

**Monochromatic spaceplates in the visible**

A nonlocal metasurface can perform space compression by implementing an angle-dependent transmission phase that matches the phase shift acquired by light propagating in free-space over a distance that exceeds the spaceplate's physical thickness. Existing spaceplate designs in the literature have been mainly based on photonic crystal slabs[12,16] or multilayered thin films made of transparent materials with relatively high refractive indices, such as Si[13,15], which are common for infrared operation, but become highly lossy and dispersive in the visible range. Among the dielectric materials with a transparency window in the visible spectrum, amorphous $TiO_2$ (refractive-index $n_{Tio2}$ ~ 2.3-2.6) and $SiO_2$ (refractive-index $n_{Sio2}$ ~ 1.46-1.48) exhibit one of the highest refractive-index contrasts available[28]. However, simply replacing Si with $TiO_2$ in existing spaceplate designs results in negligible or no space compression. To understand how to achieve good space-compression performance with lower refractive indices, we should first consider how space-compression effects are actually obtained. The most utilized physical mechanism for space compression is the angle-dependent phase and time delay provided by guided-mode resonances or FP-like resonances in planar structures. In particular, a FP cavity is arguably the simplest planar structure transmitting light at selected wavelengths with a strong angle-dependent phase shift. As detailed in Supplementary Note 1, the compression ratio ($C$) of a FP cavity made of two semi-transparent mirrors with the same reflectance ($R$), separated by a medium with a refractive index $n_c$, is as follows:

$$C = \frac{1}{n_c} \frac{1+R}{1-R} \tag{1}$$

As indicated by Eq. (1), the space compression effect of a FP structure is mainly determined by the reflectance of mirrors, with a higher reflectance offering a higher space compression ratio. We also note that the angular range and bandwidth over which this structure actually works as a spaceplate, approximating the response of free-space, inversely scale with the compression ratio and may be very narrow for a single resonator[18]. Such fundamental tradeoff can be relaxed by increasing the number of resonances, as shown in Ref.[14]. Based on these ideas, realizing a spaceplate in the visible spectrum requires the use of high-reflectance mirrors made of relatively low-index materials, with virtually zero loss, and the suitable combination and coupling of multiple resonances.

Here, we propose a design where the reflective mirrors of a planar FP cavity are implemented as dielectric distributed Bragg reflectors (DBRs) composed of $TiO_2$ and $SiO_2$ layers. The constructive interference of reflected light from multiple layers of a DBR creates high overall reflectance within a wavelength range known as stop band. For normal incidence of light at a wavelength $\lambda$, the highest reflectance occurs when each layer of the DBR is a quarter-wavelength thick, $\lambda/4n$, where $n$ is to the refractive index of the layer[29]. The number of layers in the DBR determines its reflectance and, consequently, the compression ratio of the FP cavity. Increasing the number of layers results in higher compression ratios, but reduces the usable bandwidth and angular range. Figure 1a illustrates the proposed structure of the FP cavity composed of two identical DBRs separated by a layer with half-wavelength thickness, $\lambda_r/2n_{Sio2}$, where $\lambda_r$ is the resonance wavelength. Then, as detailed in our previous work[14], multiple FP cavities with the same resonance frequencies can be coupled through quarter-wavelength spacers (Fig. 1a) to replace the desired free-space length, with high-transmission efficiency over a larger angular and frequency range.

To demonstrate the performance of the proposed structure for monochromatic space compression over the visible spectrum, we designed three spaceplates operating near the FP-like resonances at wavelengths: $\lambda_r$ = 405 nm (violet), 535 nm (green), and 700 nm (red). The DBRs implemented in each FP cavity are composed of seven alternating layers of quarter-wavelength thick $TiO_2$ and $SiO_2$, and ten FP cavities are coupled to form a spaceplate at each wavelength. Figure 1b-j shows the analytical results for the transfer functions of the three spaceplates, using the standard transfer-matrix method[30]. Figure 1b-d shows the transmission profile of the monochromatic spaceplates with respect to wavelength and incident angle for transverse-electric (TE) polarized light in free-space, where the bright lines correspond to the dispersion curves of the modes supported by the structure (Supplementary Fig. 1 also shows a similar transmission profile for the transverse-magnetic (TM) polarized incident light). We note that the width of these bright lines can be increased with the number of coupled resonances, enabling operation over wider angular and wavelength ranges, as shown in Ref.[14]. Then, in order to achieve space compression over the widest possible angular range, the operating wavelength ($\lambda_o$) of the spaceplates is chosen to be slightly off-resonance at normal incidence (at the edge of the bright line), where transmission amplitude is still high and the phase shift has an approximately quadratic response vs. incident angle, matching that of free-space. Figure 1e-g/h-j shows the transmission amplitude/phase of TE and TM polarized light with respect to the angle of incidence for spaceplates operating at $\lambda_o$ = 404.4, 533.6, and 697.6 nm. The transmission phase of spaceplates indeed matches the phase shift for propagation in free-space over a relatively wide angular range, Fig. 1h-j. For the spaceplates operating at $\lambda_o$ = 404.4, 533.6, and 697.6 nm, a space compression factor of 9.5, 5.6, and 4.8 is achieved over numerical apertures (NAs) of 0.16, 0.2, and 0.22, respectively.

The results in Fig. 1 demonstrate the high performance of the designed spaceplates, made of transparent dielectric materials, for visible wavelengths. Within the angular window of space compression, the transmission phase accurately matches the free-space propagation phase while maintaining a high transmission amplitude. Furthermore, the transverse-invariant structure of the spaceplates leads to space compression that is virtually polarization-insensitive (the transverse wave impedances for TE and TM incident waves are similar for relatively small angles, leading to similar reflection/transmission at planar boundaries). In the proposed spaceplate structure, spaceplates with shorter operating wavelengths yield higher space compression ratios for the same number of layers. The compression ratio of the spaceplate designed to operate near $\lambda_r$ = 405 nm is almost twice as high as the one near $\lambda_r$ = 700 nm. This is mainly due to frequency dispersion and the resulting higher refractive index contrast of the $TiO_2$ and $SiO_2$ layers at shorter wavelengths[29,31] in the visible spectrum, which enhances the reflectivity of DBRs. As shown in Supplementary Fig. 2, by increasing the number of layers in the DBRs to 9-layers, a compression ratio of 9.8 is attainable also near $\lambda_r$ = 700 nm. Consistent with previous studies[14,17,18], as the compression ratio increases, the operating NA and bandwidth of the spaceplate decrease, as seen in Fig. 1. These considerations will be important for the design of multi-wavelength spaceplates with achromatic response in the next section.

The performance of monochromatic spaceplates was also investigated using the finite-difference time-domain (FDTD) method (Lumerical Inc.). Figure 2 compares field intensity distributions for propagation of a TE polarized Gaussian beam in free-space and in the presence of spaceplates operating at visible wavelengths. Each of the spaceplates studied in these FDTD simulations consists of three coupled FP cavities with a structure similar to that shown Fig. 1a. As shown in Fig. 2, spaceplates realize space compression by shifting the focal point of the beam in vacuum toward the spaceplate. Consistent with our analytical results, FDTD simulations prove that monochromatic spaceplates designed at $\lambda_r$ = 405, 535, and 700 nm provide a strong space compression ratio of 9.2, 5.4, and 4.5, respectively. Notably, the intensity profile at the focal plane of the beam propagating through the spaceplate is almost identical to the corresponding profile for propagation in free-space (Fig. 2). Hence, the nonlocal angle-dependent response of the proposed spaceplates effectively reduces the focal distance of the beam without changing its properties. In other words, the entire field distribution is shifted toward the spaceplate with minimal distortion, which implies that such spaceplates can be used to reduce the length of a monochromatic imaging system without any change in its focusing and magnification power.

**Multi-color spaceplates in the visible**

The fractional bandwidth of the spaceplates designed in the previous section is on the order of 1%. Such a limited bandwidth would be a major barrier to the miniaturization of many optical devices in the visible wavelength range, such as color imaging systems. While strong compression of free-space over the full visible spectrum appears to be fundamentally difficult (and actually impossible for large values of compression ratio[18]) as discussed above, a multi-color spaceplate operating at three (or more) distinct color channels over the visible spectrum, and not over a continuous bandwidth, is a promising alternative solution to miniaturize color imaging systems without incurring in the fundamental bandwidth tradeoffs established in Ref.[18]. However, as further discussed in the following, it is highly non-trivial to design a single spaceplate with achromatic performance, in terms of transmission efficiency, compression ratio, and angular range, at different wavelengths.

Conceptually, the proposed multi-color spaceplate is based on a combination of monochromatic spaceplates, where each monochromatic element is designed to perform space compression near one specified wavelength and function as a high-transmission filter at other wavelengths. Based on this concept, we developed a multi-color spaceplate operating near $\lambda_r$ = 405, 535, and 700 nm. Figure 3a shows the wavelength- and angle-dependent transmission profile of a monochromatic spaceplate comprising five FP cavities with 7-layers of $TiO_2$ and $SiO_2$ in their DBRs, designed at $\lambda_r$ = 405 nm (Fig. 4a). Crucial for the performance of a multi-color spaceplate, this monochromatic spaceplate transmits light with high efficiencies at the two other wavelengths, $\lambda$ = 535 and 700 nm (Fig. 3a, bottom row). In order to achieve similar high-transmission performance for the monochromatic spaceplates designed at $\lambda_r$ = 535 and 700 nm, the FP cavities implemented in these spaceplates compress free-space using the second-order resonance, such that the other two wavelengths fall within transparency windows (more generally, the order of the resonance can be used as an additional degree of freedom to design multi-wavelength spaceplates). In this configuration, DBRs in FP cavities are separated by a one-wavelength layer (instead of half-wavelength) at resonance. Figure 4a shows the structure of the FP cavities used in the monochromatic spaceplates designed at $\lambda_r =$ 535 and 700 nm. The transmission profiles of these spaceplates are shown in Fig. 3b,c (bottom row).

Figure 3d-f shows the angle-dependent transmission phase of the monochromatic spaceplates and how they match the free-space phase shift at the three selected wavelengths over a certain angular range (note that the angle-dependent free-space propagation phase changes at different wavelengths). Then, in order to combine these monochromatic spaceplates to create a multi-color spaceplate with an achromatic response, one should also account for how the presence of additional structure affects the compression ratio at each wavelength. To do that, we define a compression tensor $C$, whose elements $C_{i,j}$ correspond to the compression ratios of the monochromatic spaceplate resonating at $\lambda_r = j$ nm when light with $\lambda = i$ nm wavelength passes through it. For the monochromatic spaceplates described above, the compression tensor is:

$$C = \begin{bmatrix} C_{405,405} = 9.13 & C_{535,405} = 0.67 & C_{700,405} = 0.58 \\ C_{405,535} = 0.67 & C_{535,535} = 7.22 & C_{700,535} = 0.61 \\ C_{405,700} = 0.59 & C_{535,700} = 0.67 & C_{700,700} = 5.8 \end{bmatrix}$$

While each monochromatic spaceplate performs space compression near its resonance wavelength ($C_{i,j} > 1$, for $i = j$), refraction of light in a high-transmission filter introduces a counter space-compression effect ($C_{i,j} < 1$, for $i \neq j$). This is attributed to the refractive indices of the layers being higher than the refractive index of free-space, $n_{Tio2}$ & $n_{Sio2} > 1$ (indeed, one can easily see that propagation through a non-resonant dielectric slab in free-space leads to less-than-unity compression ratio due to refraction).

A multi-color spaceplate that achromatically compresses free-space at three wavelengths is, therefore, designed using a dispersion-engineered composition of monochromatic spaceplates that achieve the same compression factor at each wavelength, taking into account the "space expansion" effect of the other cascaded spaceplates:

$$\sum_{j=405,535,700} L_j C_{405,j} = \sum_{j=405,535,700} L_j C_{535,j} = \sum_{j=405,535,700} L_j C_{700,j} \qquad (2)$$

$L_j$ is the thickness of the monochromatic spaceplate resonating at $\lambda_r = j$ nm. The overall achromatic compression ratio of the multi-color spaceplate is then simply given by any of the sums in Eq. (2) divided by the total length. The achromatic condition specified by Eq. (2) can be met by adjusting the $L_j$ factors through the number of FP cavities ($N_j$) implemented in each monochromatic spaceplate. For the case considered here, we found that the combination of $N_{405}$ = 6, $N_{535}$ = 5, $N_{700}$ = 5 elements nearly satisfies the condition for a multi-color spaceplate with an achromatic response at the selected wavelengths. The layout of this multi-color spaceplate is schematically illustrated in Fig. 4a. Figure 4b shows the wavelength- and angle-dependent transmission profile of the spaceplate across the visible spectrum. Consistent with the transmission response of the constituent monochromatic spaceplates (Fig. 3), three high-transparency bands emerge near the wavelengths at which the multi-color spaceplate is designed to perform space compression (Fig. 4b). Figures 4c,d show the transmission amplitude and phase with respect to the angle of incidence at the three operating wavelengths of $\lambda_o$ = 404.5, 534.3, and 698.8 nm. The high transmission amplitude and the excellent match between the transmission phase of the spaceplate and of free-space at the three selected wavelengths clearly demonstrate its achromatic response, achieving a compression ratio of ~ 3 over NA ~ 0.15 (Fig. 4d), close to the numerical aperture of modern smartphones (NA ~ 0.26)[13]. Figure 4e further illustrates the achromatic performance of the presented multi-color spaceplate through FDTD simulations. Comparing the propagation of a Gaussian beam in free-space and in the presence of the spaceplate at three wavelengths, the achromatic response of the designed multi-color spaceplate is evident. The focal point of the beam shifts nearly identically across the three color channels. As shown in Fig. 4e, at each wavelength, the resonance of coupled FP cavities within the corresponding monochromatic spaceplate element determines a transverse shift of the beam (see the tapered shape of the enhanced fields in different sections of the spaceplate at different wavelengths) that results in the desired space-compression effect. Figure 4f shows that the intensity profile at the focal plane of the beam propagating through the spaceplate is almost identical to the corresponding profile for propagation in free-space.

Due to the modularity of the proposed multi-color spaceplate design, a longer length of free-space can be replaced (with the same compression ratio) by simply increasing the number of FP cavities in the constituent monochromatic spaceplates. However, crucial for its achromatic response, as the number of FP cavities increases the composition ratio of $N_{405} = 6: N_{535} = 5: N_{700} = 5$ needs to be maintained if the same constituent elements are used. As an example, Supplementary Fig. 5 shows the transmission response of a multi-color spaceplate with $N_{405} = 24$, $N_{535} = 20$, $N_{700} = 20$. The presented results demonstrate again achromatic space compression across the three selected wavelengths but for a longer length of free-space.

Finally, we note that the overall achromatic compression factor of a multi-color spaceplate is lower than that of its constituent monochromatic spaceplates operating individually at each wavelength. This is due to the counter space-compression effect introduced by the refraction of light in the rest of the structure, as discussed above. A stronger overall compression effect, however, can be realized by using monochromatic spaceplates with higher individual compression ratios at their resonance wavelengths, while redesigning the stack of layers to maintain achromaticity and high transmission. In another design, we implemented three monochromatic spaceplates with compression ratios $C = 9.6$, 12.3, and 19.9 near $\lambda_r = 400$, 533, and 700 nm, respectively. The structure of the FP cavities in each of these monochromatic spaceplates is detailed in Supplementary Fig. 6. The transmission responses presented in Fig. 5a-c show that each of the monochromatic spaceplates is transparent to incident light near $\lambda = 400$, 533, and 700 nm. Figure 5d-f shows the transmission response of a multi-color spaceplate with a dispersion-engineered composition ratio of $N_{400} = 30: N_{533} = 13: N_{700} = 5$ for achromatic space compression. The angle-dependent phase shifts of the spaceplate correspond to an overall achromatic compression ratio of $C \sim 4.6$ at the three color channels, while maintaining high transmission amplitude over NA = 0.09 (Fig. 5e,f). Figure 5g demonstrates the achromatic performance of this design across the three color channels using scalar diffraction theory calculations[32].

**Discussion**

We introduced a general method for designing multi-color spaceplates that achromatically compress free-space at three wavelengths over the visible range. We stress that the compression ratios theoretically demonstrated here would be impossible to achieve over a *continuous* bandwidth covering these three colors using the considered materials. Indeed, the fundamental limit on the compression ratio of an ideal spaceplate functioning over the entire visible range is determined by the refractive index contrast of the materials within its structure[18]. For spaceplates composed of amorphous $TiO_2$ and $SiO_2$ layers, this upper limit on compression ratio over the visible spectrum is $C \sim 2.8$. Notably, this ratio is surpassed by the multi-color spaceplates presented in this work ($C \sim 3$, and 4.6). Amorphous $TiO_2$ and $SiO_2$ are among the dielectric materials most widely used in the fabrication of metasurfaces, and provide one of the highest possible refractive-index contrast for transparent materials at visible wavelengths, making them an ideal choice for the future experimental demonstration of practical visible-light spaceplates.

In the proposed designs, monochromatic spaceplates are the constituent components of a multi-color spaceplate. Similar to how individual lenses are combined to suppress chromatic aberrations in a compound lens, monochromatic spaceplates can be suitably combined to form a dispersion-engineered multi-color spaceplate with achromatic performance. The proposed method offers a high degree of flexibility in designing monochromatic spaceplates with varying compression ratios and high transmission efficiencies at any wavelength in the visible range. By increasing the number of layers in the DBR mirrors used in the FP cavities, we designed monochromatic spaceplates with compression ratios as high as $C \sim 470$ (exceeding the highest compression ratio $C = 340$ reported at a near-infrared wavelength[15]) at visible wavelengths (Supplementary Fig. 3). This is consistent with the fact that monochromatic spaceplates based on coupled FP cavities have been found to be particularly promising to approach the fundamental thickness limits of optical systems[27].

We also note that the presented method for designing multi-color and monochromatic spaceplates is general and can be applied across various wavelength ranges. For example, in the near-infrared range, using amorphous Si and $SiO_2$ layers would also result in spaceplates with strong space compression effects. As shown in Supplementary Fig. 4, we designed a monochromatic spaceplate at $\lambda_r = 1500$ nm with a compression ratio of $C = 20.7$ and a high transmission efficiency over NA = 0.13.

Other types of space compression designs such as the pancake metalens[33] and the three-lens spaceplate[34] were recently introduced. However, unlike the nonlocal angle-dependent structures considered here, these systems are not translationally invariant in the transverse direction and require a complex optical arrangement to perform moderate space compression within a highly limited NA ($C = 2$ for NA = 0.03)[34] and a very narrow field of view[33]. In addition, their bulky configuration[34] and polarization-dependent performance[33] are other limiting factors for many applications. Conversely, in this work, the non-local transversely invariant design of the spaceplates results in alignment-free and nearly polarization-insensitive space compression within a noticeably higher NA range, and the spaceplate can be placed anywhere along the optical path with no need for additional complexity in the optical setup.

Realizing space compression in the visible spectrum may pave the way for a new generation of ultra-thin optical devices, potentially making the use of complex optical systems even more widespread in our daily lives. Even a spaceplate with a compression ratio as low as two (twofold size reduction) for multiple color channels over the visible spectrum would still make a significant impact toward the miniaturization of many optical devices. AR/VR headsets, night-vision goggles, telescopes, microscopes, and cameras are among the many optical devices that could be made thin and lightweight through the use of spaceplates. More broadly, the combination of metalenses and spaceplates may lead to optical systems finally approaching the ultimate thickness limits dictated by wave physics[2,27]. Within this context, we believe that the presented approach for designing high-performance spaceplates in the visible range, based on a realistic multilayer thin-film structure made of widely used materials, will facilitate the experimental demonstration of visible-light spaceplates in the near future, and may be suitable for scalable large-area fabrication, marking a significant step toward the realization and practical use of ultra-thin optical devices for various applications.

**Acknowledgements**

We acknowledge support from Meta Materials Inc., Ideas for Innovation (i4i) Program, and the Air Force Office of Scientific Research with Grant No. FA9550-22-1-0204, through Dr. Arje Nachman.

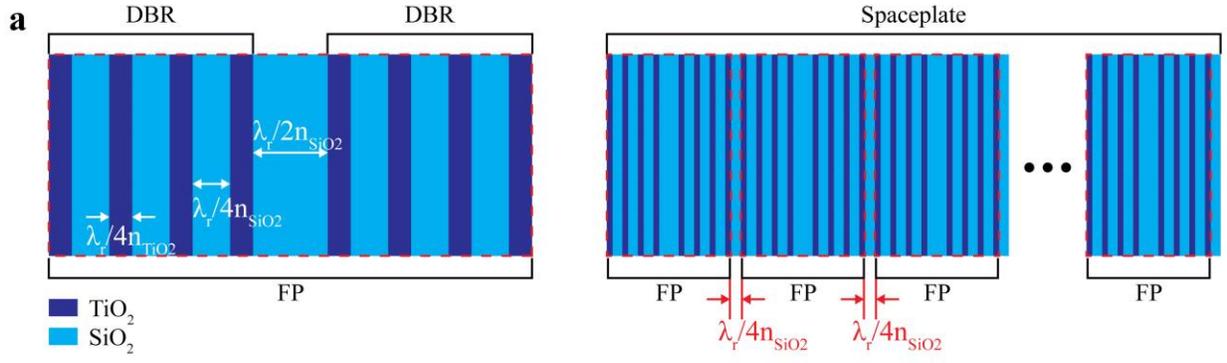

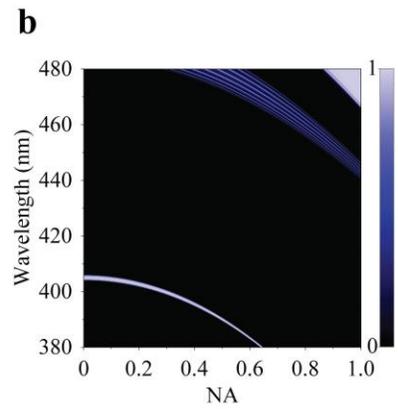
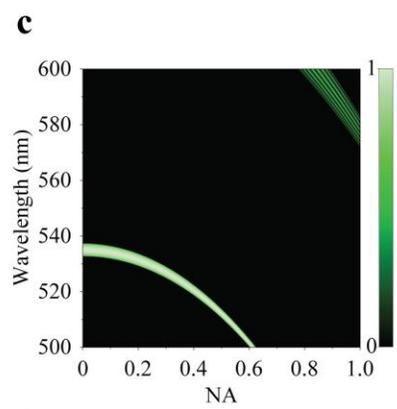
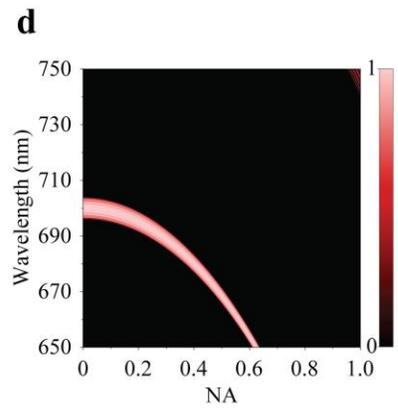

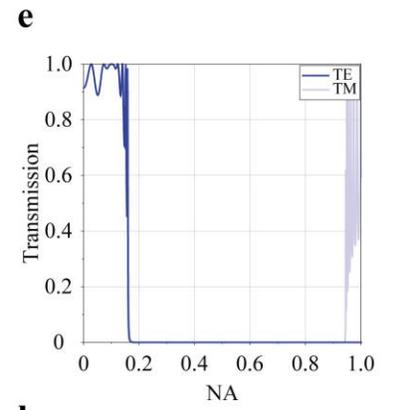
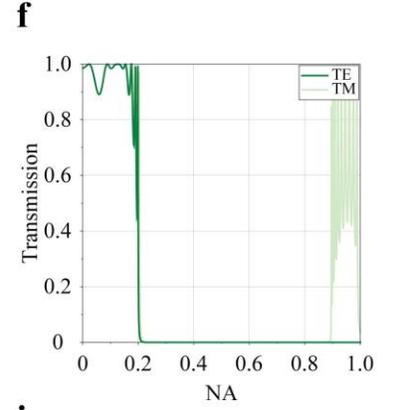
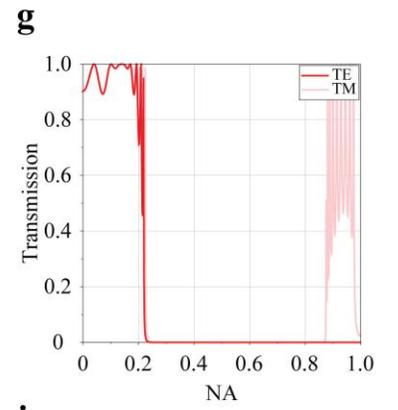

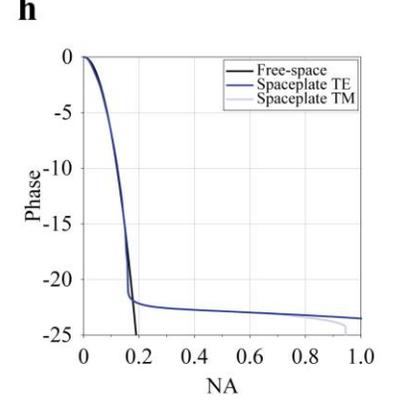
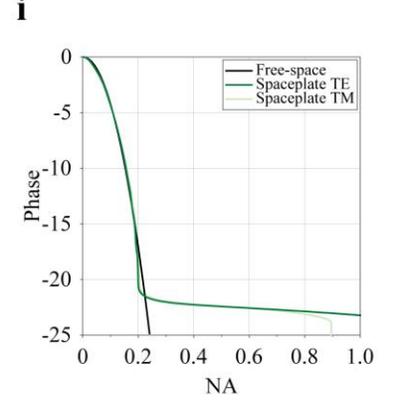
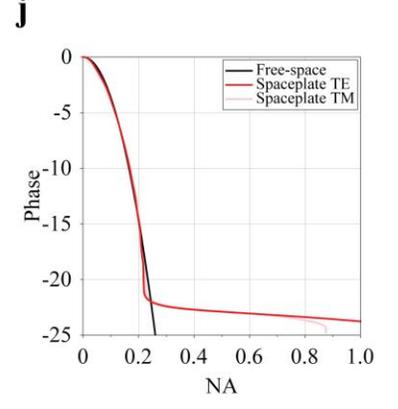

**Fig. 1 Monochromatic spaceplates in the visible. a** A spaceplate composed of coupled planar resonators. Each resonator is a planar Fabry-Perot (FP) cavity consisting of two identical DBR mirrors separated by a half-wavelength layer to achieve space compression near the resonance wavelength $\lambda_r$. Seven alternating layers of $TiO_2$ and $SiO_2$, each with a quarter-wavelength thickness, form the DBRs. The schematic on the right shows the layout of a complete spaceplate, formed by coupling identical FP cavities separated by quarter-wavelength spacers. **b-d** Transmission amplitude profiles for TE polarized light incident on spaceplates designed at resonance wavelengths of (**b**) $\lambda_r$ = 405 nm, (**c**) $\lambda_r$ = 535 nm, and (**d**) $\lambda_r$ = 700 nm. **e-g** Angle-dependent transmission amplitude for TE and TM polarized light incident on the spaceplates operating at (**e**) $\lambda_o$ = 404.4 nm, (**f**) $\lambda_o$ = 533.6 nm, and (**g**) $\lambda_o$ = 697.6 nm. **h-j** Angle-dependent transmission phase for TE and TM polarized light incident on the spaceplate compared to the phase shift acquired through free-space propagation. **h** The spaceplate designed to operate at $\lambda_o$ = 404.4 nm replaces a free-space length of ~ 88.97 μm with a thickness of ~ 9.32 μm, corresponding to a compression ratio of 9.5 over NA = 0.16. **i** The spaceplate designed to operate at $\lambda_o$ = 533.6 nm replaces a free-space length of ~ 72.04 μm with a thickness of ~ 12.73 μm, corresponding to a compression ratio of 5.6 over NA = 0.20. **j** The spaceplate designed to operate at $\lambda_o$=697.6 nm replaces a free-space length of ~ 80.22 μm with a thickness of ~ 16.87 μm, corresponding to a compression ratio of 4.8 over NA = 0.22.

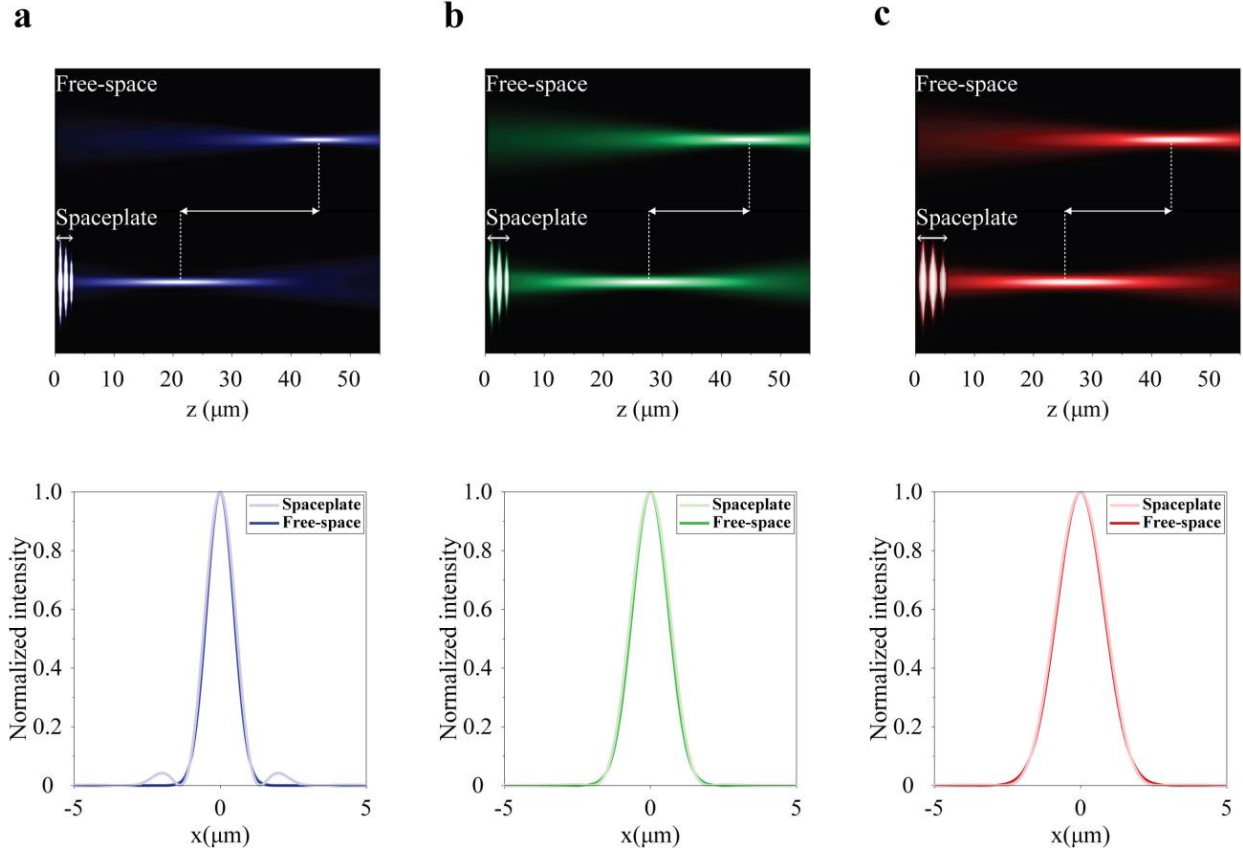

**Fig. 2. FDTD simulations of monochromatic space compression in the visible.** Normalized intensity profile for a TE polarized Gaussian beam (NA = 0.14) propagating in vacuum and in the presence of a spaceplate, calculated using the FDTD method. Each spaceplate consists of three coupled FP cavities with 7-layers of $TiO_2$ and $SiO_2$ in their DBRs. The spaceplate operated at (**a**) $\lambda_o$ = 404.4 nm with a thickness of ~ 2.79 µm, (**b**) $\lambda_o$ = 533.6 nm with a thickness of ~ 3.81 µm, and (**c**) $\lambda_o$ = 697.6 nm with a thickness of ~ 5.05 µm, moves the focal plane (**a**) 22.9 µm, (**b**) 16.7 µm, and (**c**) 17.6 µm closer to the spaceplate, which corresponds to space compression ratios of 9.2, 5.4, and 4.5, respectively. Graphs in the bottom row of **a-c** present the normalized intensity profiles across the focal planes of the beams propagating in free-space and in the presence of the designed spaceplates.

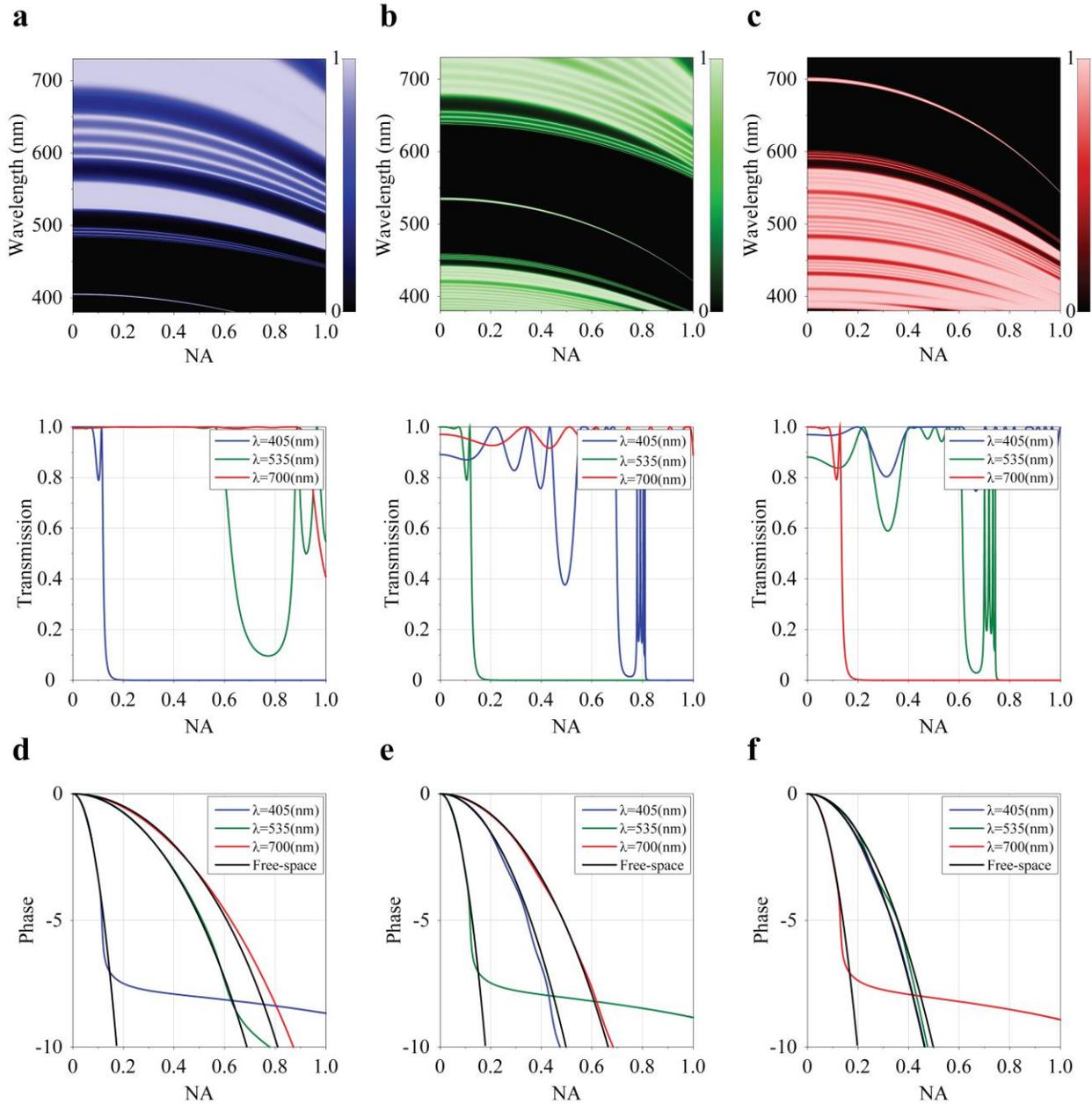

**Fig. 3. Monochromatic spaceplates as building blocks of a multi-color spaceplate. a-c** Transmission amplitude profiles for TE polarized light incident on monochromatic spaceplates consisting of five coupled FP cavities with 7-layers of $TiO_2$ and $SiO_2$ in their DBRs. These spaceplates are designed to exhibit a first-order resonance at (**a**) $\lambda_r$ = 405 nm, and a second-order resonance at (**b**) $\lambda_r$ = 535 nm and (**c**) $\lambda_r$ = 700 nm. The insets in the bottom row show the angle-dependent transmission amplitude of the spaceplates at these three wavelengths. Each spaceplate

performs space compression at one wavelength and transmits the incident light at the two other wavelengths with high efficiency. **d-f** Angle-dependent transmission phase for TE polarized light incident on the monochromatic spaceplates in (a-c) compared to the phase shift acquired through free-space propagation. **d** The monochromatic spaceplate in (a) with a thickness of ~ 4.65 μm replaces a free-space length of ~ 42.52 μm at $\lambda$ = 405 nm, ~ 3.1 μm at $\lambda$ = 535 nm, and ~ 2.7 μm at $\lambda$ = 700 nm, corresponding to compression ratios of $C_{405,405}$ = 9.13, $C_{535,405}$ = 0.67, and $C_{700,405}$ = 0.58. **e** The monochromatic spaceplate in (b) with a thickness of ~ 7.26 μm replaces a free-space length of ~ 4.86 μm at $\lambda$ = 405 nm, ~ 52.43 μm at $\lambda$ = 535 nm, and ~ 4.41 μm at $\lambda$ = 700 nm, corresponding to compression ratios of $C_{405,535}$ = 0.67, $C_{535,535}$ = 7.22, and $C_{700,535}$ = 0.61. **f** The monochromatic spaceplate in (c) with a thickness of ~ 9.62 μm replaces a free-space length of ~ 5.67 μm at $\lambda$ = 405 nm, ~ 6.42 μm at $\lambda$ = 535 nm, and ~ 56 μm at $\lambda$ = 700 nm, corresponding to compression ratios of $C_{405,700}$ = 0.59, $C_{535,700}$ = 0.67, and $C_{700,700}$ = 5.8.

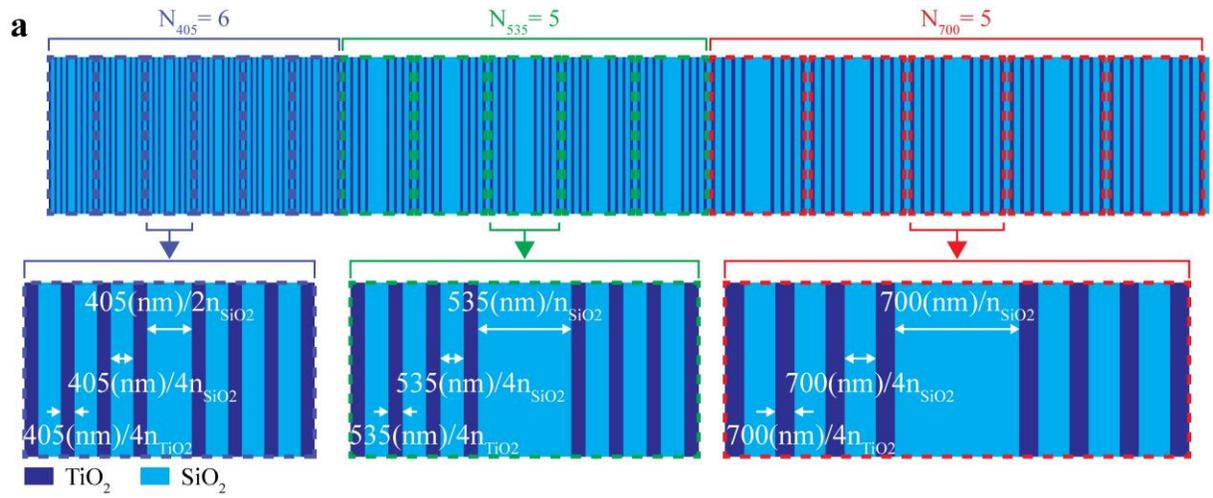
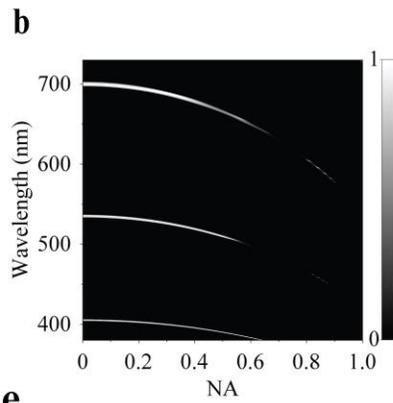
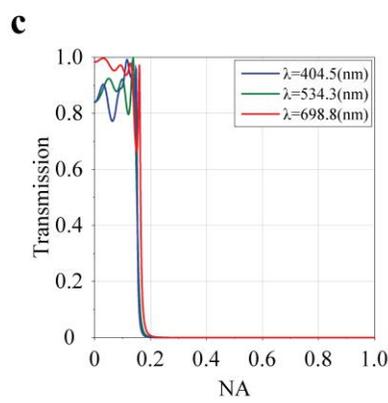
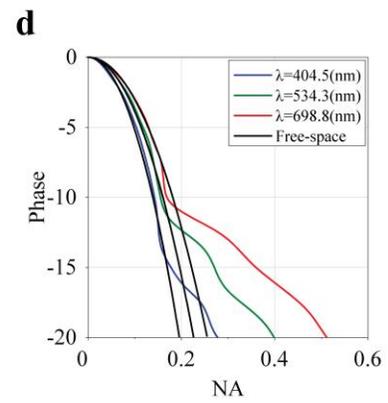
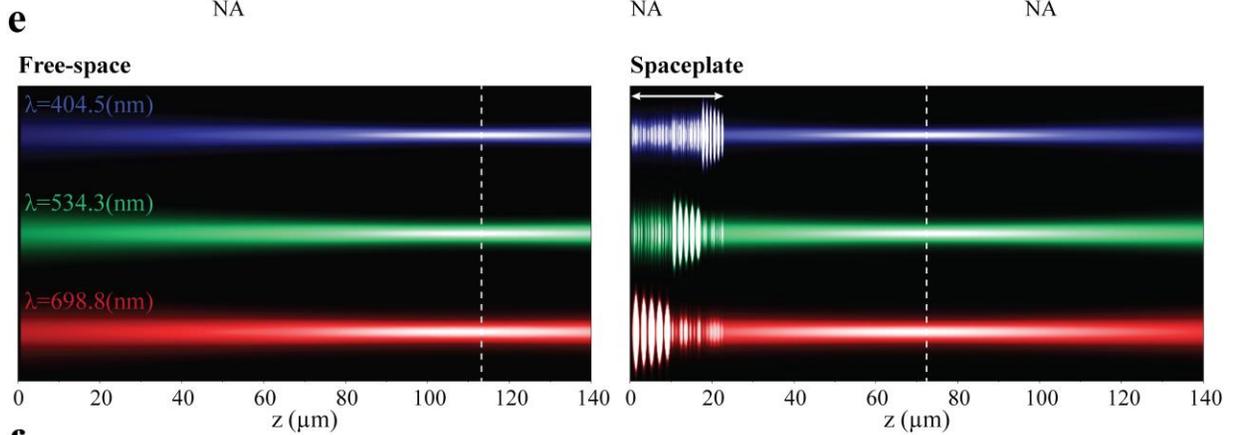
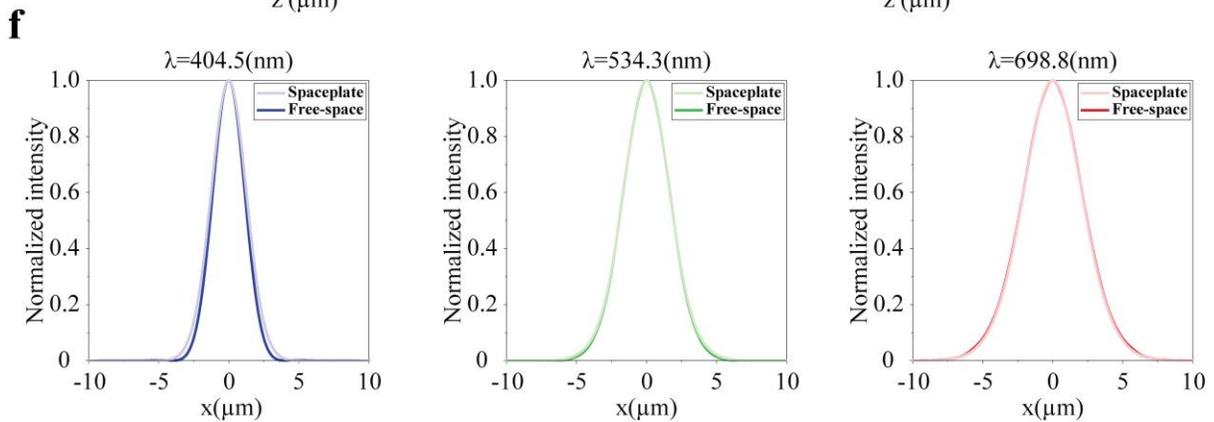

**Fig. 4. A multi-color spaceplate compressing free-space achromatically. a** Schematic of a multi-color spaceplate that performs space compression near $\lambda_r$ = 405, 535, and 700 nm, achromatically. $N_{405}$ = 6, $N_{535}$ = 5, $N_{700}$ = 5 indicate the number of FP cavities in the constituent monochromatic spaceplates resonating at $\lambda_r$ = 405, 535, and 700 nm, respectively. The insets in the bottom row show the structure of the FP cavities in each monochromatic spaceplate. **b-d** Transmission response for TE polarized light incident on the multi-color spaceplate with the structure shown in (a). **b** Transmission amplitude profile over the visible spectrum. The multi-color spaceplate is transparent near the resonance wavelengths of its constituent monochromatic spaceplates. **c** Angle-dependent transmission amplitude at the three operating wavelengths $\lambda_o$ = 404.5, 534.3, and 698.8 nm. A high transmission amplitude is maintained within a moderately wide angular range, NA = 0.15. **d** Angle-dependent transmission phase of the spaceplate, closely matching the phase response of free-space at the three operating wavelengths over NA = 0.15. The multi-color spaceplate with a thickness of ~ 22.47 µm replaces a free-space length of ~ 66.7 µm at $\lambda$ = 404.5 nm, ~ 65.7 µm at $\lambda$ = 534.3 nm, and ~ 67.1 µm at $\lambda$ = 698.8 nm, corresponding to almost identical compression ratios of $C$ ~ 3.0, 2.9, and 3.0, respectively. These results highlight the achromatic performance of the multi-color spaceplate at these three wavelengths. **e** Normalized intensity profile for a TE polarized Gaussian beam (NA = 0.05) propagating in vacuum and in the presence of the multi-color spaceplate in (a), calculated at the three operating wavelengths using the FDTD method. Compared to free-space propagation, the focal plane shifts by ~ 42 µm at $\lambda$ = 404.5 nm, ~ 39.8 µm at $\lambda$ = 534.3 nm, and ~ 40.7 µm at $\lambda$ = 698.8 nm toward the spaceplate. **f** Normalized intensity profiles across the focal planes of the beams in (e).

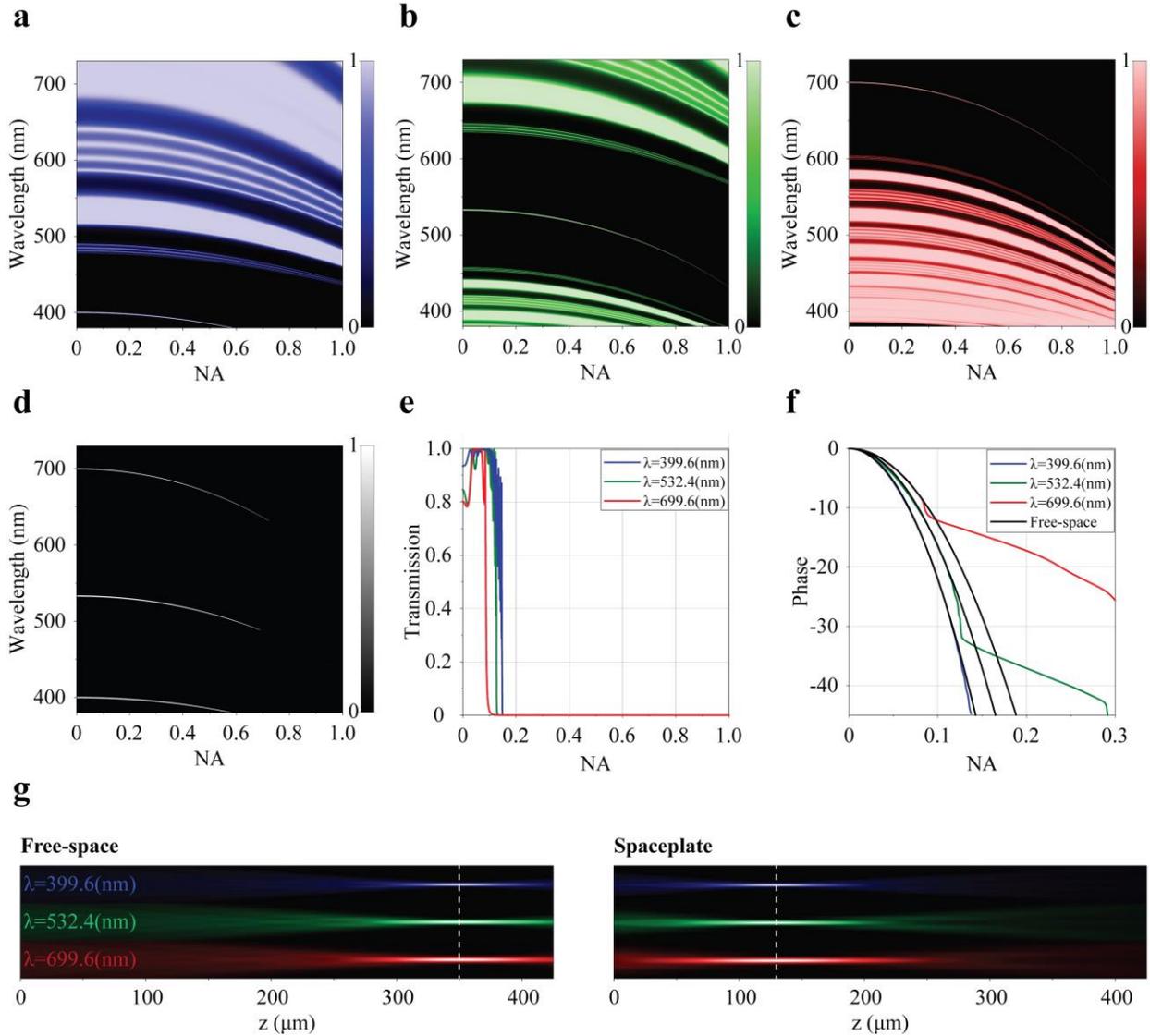

**Fig. 5. A multi-color spaceplate with a high compression ratio. a-c** Transmission amplitude profiles for TE incident light on monochromatic spaceplates designed at (**a**) $\lambda_r$ = 400 nm (**b**) $\lambda_r$ = 533 nm (**c**) $\lambda_r$ = 700 nm over the visible spectrum. The spaceplates consist of five FP cavities with (**a**) seven (**b**) nine (**c**) eleven alternating layers of $TiO_2$ and $SiO_2$ in their DBRs (Supplementary Fig. 6). **d** Transmission amplitude profile for TE incident light on a multi-color spaceplate with $N_{400}$ = 30, $N_{533}$ = 13, $N_{700}$ = 5 composition (Supplementary Fig. 6). **e,f** Angle-dependent transmission amplitude (**e**) and phase (**f**) of the multi-color spaceplate at $\lambda_o$ = 399.6,

532.4, and 699.6 nm. This spaceplate with a thickness of ~ 60.1 μm replaces a free-space length of ~ 279.7 μm at $\lambda$ = 399.6 nm, ~ 278 μm at $\lambda$ = 532.4 nm, and ~ 279.8 μm at $\lambda$ = 699.6 nm, corresponding to an achromatic overall compression ratio of $C$ ~ 4.6 at the three wavelengths. **g** Normalized intensity profile for a TE polarized Gaussian beam (NA = 0.09) propagating in vacuum and in the presence of the multi-color spaceplate, calculated at the three selected wavelengths using scalar diffraction theory. The spaceplate shows an achromatic response at these three wavelengths (221, 224, and 219.5 μm shift at $\lambda$ = 399.6, 532.4, and 699.6 nm, respectively).